\documentclass[10pt]{article}
\usepackage{amsmath}
\usepackage{amsthm}
\usepackage{amssymb}
\usepackage{graphicx}

\makeatletter

\def\catchNameTitle{
\global\let\Author\@author \global\let\Title\@title }

\def\timenow{%
  \@tempcnta=\time \divide\@tempcnta by 60 \number\@tempcnta:\multiply
  \@tempcnta by 60 \@tempcntb=\time \advance\@tempcntb by -\@tempcnta
  \ifnum\@tempcntb <10 0\number\@tempcntb\else\number\@tempcntb\fi}

\def\TODAY{\number\year-\ifcase\month\or 01\or 02\or 03\or 04\or 05\or
    06\or 07\or 08\or 09\or 10\or 11\or 12\fi-\ifnum\day<10 0\number\day\else \number\day\fi}

\newcommand{\MarkAll}[4]{
\renewcommand{\ps@myheadings}{
\renewcommand{\@oddhead}{{\scriptsize #1 \hfil #2}}
\renewcommand{\@evenhead}{\@oddhead}
\renewcommand{\@evenfoot}{{\scriptsize #3}\hfil
    \textrm{\thepage}\hfil{\scriptsize #4}}
\renewcommand{\@oddfoot}{\@evenfoot}}
}

\def\NOW{\TODAY;\timenow}

\newcommand{\DRAFT}{
\renewcommand{\@evenhead}{\hfil DRAFT \hfil}
\renewcommand{\@oddhead}{\@evenhead}
}

\newcommand{\NowFootNum}{
\renewcommand{\@evenfoot}{{\scriptsize \jobname.tex}\hfil
    \textrm{\thepage}\hfil {\scriptsize \NOW}}
\renewcommand{\@oddfoot}{\@evenfoot}
}

\newcommand{\NowFoot}{
\renewcommand{\@evenfoot}{{\scriptsize \jobname.tex} {\scriptsize \NOW}}
\renewcommand{\@oddfoot}{\@evenfoot}
}

\def\fixNumberingInArticle{
\@addtoreset{figure}{section}
\renewcommand{\thefigure}{\thesection.\arabic{figure}}
\renewcommand{\theequation}{\arabic{equation}}
}

\makeatother

\newcommand{\lb}{\left(}
\newcommand{\rb}{\right)}

\newtheorem{lemma}{Lemma}
\newtheorem{prop}{Proposition}

\newtheorem{thm}{Theorem}

\newcommand{\dx}{\partial_x}
\newcommand{\dt}{\partial_t}
\newcommand{\dc}{\partial_c}
\newcommand{\p}{\partial}
\newcommand{\dd}[1]{\frac{d}{d #1}}

\newcommand{\SympOp}{\dx^{-1}}
\newcommand{\SympOpInv}{\dx}
\newcommand{\RestSympOpInv}{{\Omega_{c a}^{-1}}}
\newcommand{\RestSympOp}{{\Omega_{c a}}}

\newcommand{\intR}[1]{\int_{-\infty}^\infty #1\, dx}

\newcommand{\TMProj}{P_T}

\newcommand{\Null}[1]{\mbox{\rm Null}\ #1}

\newcommand{\Span}[1]{ \mbox{\rm Span}\ \left\{#1\right\}}

\newcommand{\HamiltonianWithPotential}{H_\B}
\newcommand{\HamiltonianWithoutPotential}{H_{\B=0}}

\newcommand{\Lca}{\mathcal{L}_{Q}}

\newcommand{\ActionAtSoliton}{\Lambda_{ca}}
\newcommand{\LyapunovFunctional}{\Gamma_c}

\renewcommand{\O}[1]{\mbox{\rm O}\left( #1\right)}

\newcommand{\Qc}{Q_c}
\newcommand{\Qca}{Q_{ca}}

\newcommand{\va}{\zeta^{tr}_{c a}}
\newcommand{\vc}{\zeta^n_{c a}}

\newcommand{\e}{\xi}

\newcommand{\DB}{\delta \B}
\newcommand{\RB}{\delta^2\B}
\newcommand{\ev}{\epsilon_x}
\newcommand{\av}{\epsilon_a}
\newcommand{\tv}{\epsilon_t}
\newcommand{\B}{b}

\newcommand{\NpA}[1]{N'_{c a}(#1)}
\newcommand{\NA}[1]{N_{c a}(#1)} 

\newcommand{\R}{{\mathbb R}}
\newcommand{\Rp}{\R_+}

\newcommand{\Lp}[1]{L^{#1}(\R)}
\newcommand{\Hs}[1]{H^{#1}(\R)}

\newcommand{\TcaM}{T_{\Qca}M_s}

\newcommand{\LpNorm}[2]{\|#2\|_{L^{#1}}}
\newcommand{\HsNorm}[2]{\|#2\|_{H^{#1}}}
\newcommand{\HTNorm}[1]{\left\|#1\right\|_{T}}
\newcommand{\SupNorm}[1]{\|#1\|_\infty}
\newcommand{\ip}[2]{\langle #1,#2\rangle}
\newcommand{\InfI}[1]{\lfloor #1\rfloor}

\title{Long-Time Dynamics of Variable Coefficient mKdV Solitary Waves\thanks{This paper is part of the first author's Ph.D. thesis.}}
\author{S.I. Dejak$^{\S \natural\ddagger}$\thanks{Supported by NSERC under grant NA7901 and Ontario Graduate Scholarships.}  and
B.L.G.
Jonsson$^{\flat \sharp}$\thanks{Supported by the Swiss National Foundation under NF-Project 20-105493 and by "The Year of PDEs" at the Fields Institute.} \\
\small $^\S$University of Notre Dame, Notre Dame, U.S.A.\\\small
$^\natural$University of Toronto, Toronto, Canada\\
\small $^\flat$Theoretische Physik, ETH-H\"{o}nggerberg, CH-8039
Z\"{u}rich, Switzerland\\
\small $^\sharp$The Alf\'{e}nlaboratory, Royal Institute of
Technology, 100 44 Stockholm, Sweden}
\date{}
\begin{document}
\maketitle \fixNumberingInArticle

{\abstract We study the Korteweg-de Vries-type equation $\dt
u=-\dx\lb\dx^2 u+f(u)-\B(t,x)u\rb$, where $\B$ is a small and
bounded, slowly varying function and $f$ is a nonlinearity. Many
variable coefficient KdV-type equations can be rescaled into this
equation. We study the long time behaviour of solutions with
initial conditions close to a stable, $\B=0$ solitary wave.  We
prove that for long time intervals, such solutions have the form
of the solitary wave, whose centre and scale evolve according to a
certain dynamical law involving the function $\B(t,x)$, plus an
$\Hs{1}$-small fluctuation.}

\section{Introduction}
\indent\indent We study the long time behaviour of solutions to a
class Korteweg-de Vries-type equations, with an additional term
$\B(t,x)u$.  These equations, from now on called the bKdV, are of
the form
\begin{align}
 \dt u=-\dx\lb\dx^2 u+f(u)-\B(t,x)u\rb,
\label{Eqn:KdvGeneralizedWithPotential}
\end{align}
where $\B(t,x)$ is a real valued function and $f$ is a
nonlinearity. In this paper we consider a restricted class of
nonlinearities. In particular, for monomial nonlinearities, we
give a result only for $f(u)=u^3$, corresponding to the modified
KdV (mKdV).  When $\B=0$, Equation
\eqref{Eqn:KdvGeneralizedWithPotential} reduces to the generalized
Korteweg-de Vries equation (GKdV)
\begin{equation}
\dt u=-\dx(\dx^2 u+f(u)). \label{Eqn:GKdV}
\end{equation}

A remarkable property of the GKdV is the existence of spatially
localized solitary (or travelling) waves, i.e. solutions of the
form $u=\Qc(x-a-c t)$, where $a\in\R$ and $c$ in some interval
$I$. When $f(u)=u^p$ and $p\ge 2$, solitary waves are explicitly
computed to be
\begin{equation*}
\Qc(x)=c^\frac{1}{p-1}Q(c^\frac{1}{2}x),
\end{equation*}
where
\begin{equation*}
Q(x)=\lb\frac{p+1}{2}\rb^\frac{1}{p-1} \lb\cosh\lb\frac{p-1}{2}
x\rb\rb^2.
\end{equation*}
It is generally believed that an arbitrary, say $\Hs{1}$, solution
to equation \eqref{Eqn:GKdV} eventually breaks up into a
collection of solitary waves and radiation.  A discussion of this
phenomenon for the generalized KdV appears in Bona \cite{BoSo94}.
 For the general, but integrable case see Deift and Zhou \cite{DeZh93}.

The mKdV equation is fundamental in many areas of applied
mathematics ranging from traffic flow to plasma physics (see
\cite{Ku1985,ChRa1987,LuMa1997,Na2002}) and arises from an
approximation of a more complicated systems. The effects of higher
order processes can often be collected into a term of the form
$\B(t,x) u$.  Our main result stated at the end of the next
section gives, for long time, an explicit, leading order
description of a solution to the bKdV initially close to a
solitary wave solution of the GKdV.

We assume that the coefficient $\B$ and nonlinearity $f$ are such
that \eqref{Eqn:KdvGeneralizedWithPotential} has global solutions
for $\Hs{1}$ data and that \eqref{Eqn:KdvGeneralizedWithPotential}
with $\B=0$ possesses solitary wave solutions.  Precise conditions
will be formulated in the next section.  Here we mention that the
literature regarding well-posedness of the KdV ($\B=0$,
$f(u)=u^2$) is extensive and well developed. The Miura transform
(see \cite{Mi1968}) then gives well-posedness results for the
mKdV. Bona and Smith \cite{BoSm1975} proved global wellposedness
of the KdV in $\Hs{2}$.  See also \cite{Ka1983}. Kenig, Ponce, and
Vega \cite{KePoVe1996} have proved local wellposedness in $\Hs{s}$
for $s\ge -\frac{3}{4}$ and similar results are available for the
generalized KdV ($\B=0$, monomial nonlinearity $f(u)=u^p$ with
$p=2,3,4$)\cite{KePoVe1993}. In particular, local well-posedness
for the mKdV in $\Hs{s}$ with $s\ge\frac{1}{4}$ and global
well-posedness for $s\ge 1$ are known.  More recently, results
extending local wellposedness in negative index Sobolev spaces to
global wellposedness have been proven \cite{CoSt1999,CoKe2001}.
There is little literature on global well-posedness of the bKdV in
energy space, however, under a smallness assumption on the
coefficient $\B$, Dejak and Sigal \cite{DeSi2004} proved global
well-posedness in $\Hs{1}$ of the bKdV with $f(u)=u^p$, $p=2,3,4$.
They used results of \cite{KePoVe1993}, and perturbation and
energy arguments.

Soliton solutions of the KdV equation are known to be orbitally
stable.  Although the linearized analysis of Jeffrey and Kakutani
\cite{JeKa1970} suggested orbital stability, the first nonlinear
stability result was given by Benjamin \cite{Be1972}.  He assumed
smooth solutions and used Lyapunov stability and spectral theory
to prove his results.  Bona \cite{Bo1975} later corrected and
improved Benjamin's result to solutions in $\Hs{2}$. Weinstein
\cite{We1985} used variation methods, avoiding the use of an
explicit spectral respresentaion, and extended the orbital
stability result to the GKdV.  More recently, Grillakis, Satah,
and Strauss \cite{GrShSt87} extended the Lyapunov method to
abstract Hamiltonian systems with symmetry.  Numerical simulations
of soliton dynamics for the KdV were performed by Bona et al. See
\cite{BoDo86,BoDo91,BoDo95,BoDo96}.

For nonlinear Schr\"{o}dinger and Hartree equations, long-time
dynamics of solitary waves were studied by Bronski and Gerrard
\cite{BrJe2000}, Fr\"{o}hlich, Tsai and Yau \cite{FrTs2003},
Keraani \cite{Ke02}, and Fr\"{o}hlich, Gustafson, Jonsson, and
Sigal \cite{FrGu2003, FrGuJoSi2005}.  For related results and
techniques for the nonlinear Schr\"{o}dinger equations see also
\cite{BuPe1992,BuSu2003,GaSi2004,RoScSoPreprint,RoScSoPreprintII,TsYa2002I,TsYa2002II,TsYa2002III,SoWe90}.

In our approach we use the fact that the bKdV is a
(non-autonomous, if $b$ depends on time) Hamiltonian system.  As
in the case of the nonlinear Schr\"{o}dinger equation (see
\cite{FrGu2003}), we construct a Hamiltonian reduction of this
original, infinite dimensional dynamical system to a two
dimensional dynamical system on a manifold of soliton
configurations.  The analysis of the general bKdV immediately runs
into the problem that the natural symplectic form $\omega$ is not
defined on the tangent space of the soliton manifold.  In this
paper we prove the main theorem in the cases where the symplectic
form is well defined on the tangent space.  One such case is when
the nonlinearity is $f(u)=u^3$.  For the general case see
\cite{DeSi2004}.  We remark here that the dynamics for the special
case considered here include the higher order correction terms for
the scaling parameter $c$, which cannot be included in the general
case.

\vspace{4mm}\noindent{\bf\large Acknowledgements} \vspace{4mm}\\
We are grateful to I.M. Sigal for useful discussions.

\section{Preliminaries, Assumptions, Main Results}
\label{Section:AssumptionsAndMainResults}The bKdV can be written
in Hamiltonian form as
\begin{eqnarray}
\dt u=\SympOpInv\HamiltonianWithPotential'(u),
\label{Eqn:KdVVariationalForm}
\end{eqnarray}
where $\HamiltonianWithPotential'$ is the $\Lp{2}$ function
corresponding to the Fr\'{e}chet derivative
$\p\HamiltonianWithPotential$ in the $\Lp{2}$ pairing.  Here the
Hamiltonian $\HamiltonianWithPotential$ is
\begin{eqnarray*}
\HamiltonianWithPotential(u):=\intR{ \frac{1}{2}(\dx
u)^2-F(u)+\frac{1}{2} \B(t,x) u^2},
\end{eqnarray*}
where the function $F$ is the antiderivative of $f$ with $F(0)=0$.
The operator $\dx$ is the anti-self-adjoint operator (symplectic
operator) generating the Poisson bracket
\begin{equation*}
\{G_1, G_2\}:=\frac{1}{2}\intR{G_1'(u)\dx G_2'(u)-G_2'(u)\dx
G_1'(u)},
\end{equation*}
defined for any $G_1$, $G_2$ such that $G_1', G_2'\in
\Hs{\frac{1}{2}}$.
 The corresponding symplectic form is
\begin{equation*}
\omega(v_1,\, v_2):=\frac{1}{2}\intR{v_1(x)\dx^{-1}
v_2(x)-v_2(x)\dx^{-1} v_1(x)},
\end{equation*}
defined for any $v_1, v_2\in \Lp{1}$.  Here the operator
$\dx^{-1}$ is defined as
\begin{equation*}
\dx^{-1} v(x):=\int_{-\infty}^x v(y)\, dy.
\end{equation*}
Note that $\dx^{-1}\cdot \dx=I$ and, on the space
$\{u\in\Lp{2}\,|\, \intR{u}=0\}$, $\dx^{-1}$ is formally
anti-self-adjoint with inverse $\dx$.  Hence, if
$\intR{v_1(x)}=0$, then $\omega(v_1,\, v_2)=\intR{v_1(x)\dx^{-1}
v_2(x)}$.

Note that if $\B$ depends on time $t$, then equation
\eqref{Eqn:KdVVariationalForm} is non-autonomous.  It is, however,
in the form of a conservation law, and hence the integral of the
solution $u$ is conserved provided $u$ and its derivatives decay
to zero at infinity:
\begin{equation*}
\dd{t}\intR{u}=0.
\end{equation*}
There are also conserved quantities associated to symmetries of
\eqref{Eqn:KdvGeneralizedWithPotential} when $\B=0$. The simplest
such corresponds to time translation invariance and is the
Hamiltonian itself.  This is also true if $\B$ is non-zero but
time independent. If the potential $\B=0$, then
\eqref{Eqn:KdvGeneralizedWithPotential} is also spatially
translation invariant. Noether's theorem then implies that the
flow preserves the momentum
\begin{eqnarray*}
P(u):=\frac{1}{2}\LpNorm{2}{u}^2.
\end{eqnarray*}
In general, when $\B\ne 0$ the temporal and spatial translation
symmetries are broken, and hence, the Hamiltonian and momentum are
no longer conserved. Instead, one has the relations
\begin{align}
\dd{t}\HamiltonianWithPotential(u)&=\frac{1}{2}\intR{(\dt\B) u^2},
\label{Eqn:ConservationHamiltonian}\\
\dd{t}P(u)&=\frac{1}{2}\intR{\B'u^2},
\label{Eqn:ConservationMomentum}
\end{align}
where $\B'(t,x):=\dx \B(t,x)$.  For later use, we also state the
relation
\begin{eqnarray}
\dd{t}\frac{1}{2}\intR{\B u^2}=\intR{\frac{1}{2} u^2\dt\B+\B'\lb u
f(u)-\frac{3}{2}(\dx u)^2-F(u)\rb-\B'' u \dx u}.
\label{Eqn:ConservationPotentialMomentum}
\end{eqnarray}
Assuming (\ref{Eqn:KdvGeneralizedWithPotential}) is well-posed in
$\Hs{2}$, the above equalities are obtained after multiple
integration by parts.  Then, by density of $\Hs{2}$ in $\Hs{1}$,
the equalities continue to hold for solutions in $\Hs{1}$.  To
avoid these technical details, we assume the Hamiltonian flow on
$\Hs{1}$ enjoys (\ref{Eqn:ConservationHamiltonian}),
(\ref{Eqn:ConservationMomentum}) and
(\ref{Eqn:ConservationPotentialMomentum}).

Consider the GKdV, i.e. equation \eqref{Eqn:GKdV}.  Under certain
conditions on $f$, this equation has travelling wave solutions of
the form $\Qc(x-c t)$, where $\Qc$ a positive $\Hs{2}$ function.
Substituting $u=\Qc(x-ct)$ into the GKdV gives the scalar field
equation
\begin{equation}
-\dx^2\Qc+c\Qc-f(\Qc)=0. \label{Eqn:ScalarFieldEquation}
\end{equation}
Existence of solutions to this equation has been studied by
numerous authors. See \cite{St1977, BeLi1983}.  In particular,
Berestyki and Lions \cite{BeLi1983} give sufficient and necessary
conditions for a positive and smooth solution $\Qc$ to exist.  We
assume $g:=-c u+f(u)$ satisfies the following conditions:
\begin{enumerate}
\item $g$ is locally Lipschitz and $g(0)=0$, \item
$x^*:=\inf\{x>0\,|\,\int_0^x g(y)\,dy\}$ exists with $x^*>0$ and
$g(x^*)>0$, and \item $\lim_{s\rightarrow 0}\frac{g(s)}{s}\le
-m<0$.
\end{enumerate}
Then, as shown by Berestycki and Lions,
\eqref{Eqn:ScalarFieldEquation} has a unique (modulo translations)
solution $\Qc\in C^2$ for $c$ in some interval, which is positive,
even (when centred at the origin), and with $\Qc$, $\dx\Qc$, and
$\dx^2\Qc$ exponentially decaying to zero at infinity ($\dx\Qc<0$
for $x>0$). Furthermore, if $f$ is $C^2$, then the implicit
function theorem implies that $\Qc$ is $C^2$ with respect to the
parameter $c$ on some interval $I_0\subset\Rp$.  We assume that
$x^m\dc^n\Qc\in\Lp{1}$ for $n=1,2$, $m=0,1,2$ so that integrals
containing $\dc^n\Qc$ are continuous and differentiable with
respect to $c$.
 We also make the assumption that
\begin{equation}
\intR{\dc\Qc}=0 \label{Eqn:IntVcAssump}
\end{equation}
for all $c\in I$.  This implies that
\begin{eqnarray}
\int_{-\infty}^x \dc\Qc(z)\, dz, \int_{-\infty}^x \dc^2\Qc(z)\,
dz\in\Lp{2}. \label{Eqn:LIIAssumptions}
\end{eqnarray}
To see this use the isometry property of the Fourier transform and
the decay properties of $\dc\Qc$. The above requirements of $\Qc$
are implicit assumptions on the nonlinearity $f$ and are true when
$f(u)=u^3$. Assumption \eqref{Eqn:IntVcAssump} is a very important
and restrictive requirement; it does not hold when $f(x)=x^p$ and
$p\ne 3$.  For the case where \eqref{Eqn:IntVcAssump} does not
hold see \cite{DeSi2004}.

The solitary waves $\Qc$ are orbitally stable if $\delta'(c)>0$,
where $\delta(c)=P(\Qc)$.  See Weinstein \cite{We1985} the first
proof for general nonlinearities. Moreover, in \cite{GrShSt87},
Grillakis, Shatah and Strauss proved that $\delta'(c)>0$ is a
necessary and sufficient condition for $\Qc$ to be orbitally
stable.  In this paper, we assume that $\Qc$ is stable for all $c$
in some compact interval $I\subset I_0$, or equivalently that
$\delta'(c)>0$ on $I$.  For $f(u)=u^p$, we have
$\delta'(c)=\frac{5-p}{4(p-1)}\LpNorm{2}{Q_{c=1}}^2$, which
implies the well known stability criterion $p<5$ corresponding to
subcritical power nonlinearities.

The scalar field equation \eqref{Eqn:ScalarFieldEquation} for the
solitary wave can be viewed as an Euler-Lagrange equation for the
extremals of the Hamiltonian $\HamiltonianWithoutPotential$
subject to constant momentum $P(u)$.  Moreover, $\Qc$ is a stable
solitary wave if and only if it is a minimizer of
$\HamiltonianWithoutPotential$ subject to constant momentum $P$.
Thus, if $c$ is the Lagrange multiplier associated to the momentum
constraint, then $\Qc$ is an extremal of
\begin{align}
\ActionAtSoliton(u)&:=\HamiltonianWithoutPotential(u)+c P(u)
\label{DefinitionLS}\\&=\intR{ \frac{1}{2}(\dx u )^2+\frac{1}{2}c
u^2-F(u)},\nonumber
\end{align}
and hence $\ActionAtSoliton'(\Qc)=0$.

The functional $\ActionAtSoliton$ is translationally invariant.
Therefore, $\Qca(x):=\Qc(x-a)$ is also an extremal of
$\ActionAtSoliton$, and $\Qc(x-c t-a)$ is a solitary wave solution
of (\ref{Eqn:KdvGeneralizedWithPotential}) with $\B=0$. All such
solutions form the two dimensional $C^\infty$ manifold of solitary
waves
\begin{equation*}
M_s:=\{\Qca\,|\,c\in I, a\in \R\},
\end{equation*}
with tangent space $\TcaM$ spanned by the vectors
\begin{eqnarray}
\va:=\p_a\Qca=-\dx\Qca\ \mbox{and}\ \vc:=\p_c\Qca,
\label{Eqn:DefinitionOfTangentVectors}
\end{eqnarray}
which we call the translation and normalization vectors.  Notice
that the two tangent vectors are orthogonal in $\Lp{2}$.

In addition to the requirement on $\B$ that
(\ref{Eqn:KdvGeneralizedWithPotential}) is globally wellposed, we
assume the potential $\B$ is bounded, twice differentiable, and
small in the sense that
\begin{align}
|\dt^n\dx^m\B|\le\av\tv^n\ev^m, \label{Eqn:AssumptionOnPotential}
\end{align}
for $n=0,1$, $m=0,1,2$, and $n+m\le 2$.  The positive constants
$\av$, $\ev$, and $\tv$ are amplitude, length, and time scales of
the function $\B$. We assume all are less than or equal to one.

Lastly, we make some explicit assumptions on the local
nonlinearity $f$. We require the nonlinearity to be $k$ times
continuously differentiable, with $f^{(k)}$ bounded for some $k\ge
3$ and $f(0)=f'(0)=0$.  These assumptions ensure the Hamiltonian
is finite on the space $\Hs{1}$ and, since $\Qc$ decays
exponentially (see \cite{BeLi1983}), both $f(\Qc)$ and $f'(\Qc)$
have exponential decay.

We are ready to state our main result.  Recall that $I_0\subset
\R_+$ is an interval where $\Qc$ is twice continuously
differentiable.
\begin{thm}
Let the above assumptions hold and assume $\delta'(c)>0$ for all
$c$ in a compact set $I\subset I_0$.  Assume $\av\le 1$.  Then, if
$\ev\le 1$, $\epsilon_0$ and $\tv$ are small enough, there is a
positive constant $C$ such that the solution to
(\ref{Eqn:KdvGeneralizedWithPotential}) with an initial condition
$u_0$ satisfying $\inf_{\Qca\in M_s}\HsNorm{1}{u_0-Q_{c a}}\le
\epsilon_0$ can be written as
\begin{eqnarray*}
u(x,t)=Q_{c(t)}(x-a(t))+\e(x,t),
\end{eqnarray*}
where $\HsNorm{1}{\e(t)}=\O{
\epsilon_0+(\av\ev\epsilon_0)^\frac{1}{2}+\ev+\tv}$ for all times
$t \le C(\av \ev)^{-1}$.  Moreover, during this time interval the
parameters $a(t)$ and $c(t)$ satisfy the equations
\begin{eqnarray*}
\lb\begin{array}{c}
  \dot{a} \\
  \dot{c}
\end{array}\rb&=&
\lb\begin{array}{c}
c-\B(a)\\
0
\end{array}\rb+
\B'(a)\frac{\delta(c)}{\delta'(c)}\lb\begin{array}{c}
0\\
1
\end{array}\rb+\O{(\epsilon_0+\ev+\tv)^2+(\av\ev\epsilon_0)^\frac{1}{2}(\ev+\tv+\epsilon_0)},
\end{eqnarray*}
where $c$ is assumed to lie in the compact set $I$.
\label{MainThm}
\end{thm}
\begin{proof}[Sketch of Proof and Paper Organization]
To realize the Hamiltonian reduction we decompose functions in a
neighbourhood of the soliton manifold $M_s$ as
\begin{equation*}
u=\Qca+\e
\end{equation*}
with $\e$ symplectically orthogonal to $\TcaM$, i.e.
$\e\bot\dx^{-1}\TcaM$.  We show that there is an $\epsilon_0>0$
such that if the solution $u$ satisifes the estimate
$\inf_{\Qca}\HsNorm{1}{u-\Qca}<\epsilon_0$, then there are unique
$C^1$ functions $a(u)$ and $c(u)$ such that $u=Q_{c(u) a(u)}+\e$
with $\e\bot\dx^{-1}\TcaM$.

With the knowledge that the symplectic decomposition exists, we
substitute $u=\Qca+\e$ into the bKdV
\eqref{Eqn:KdvGeneralizedWithPotential} and split the resulting
equation according to the decomposition
\begin{equation*}
\Lp{2}=\dx^{-1}\TcaM\oplus\lb\dx^{-1}\TcaM\rb^{\bot}
\end{equation*}
to obtain equations for the parameters $c$ and $a$, and an
equation for the (infinite dimensional) fluctuation $\e$.  In
Section \ref{Section:Projection} we isolate the leading order
terms in the equations for $a$ and $c$ and estimate the remainder,
including all terms containing $\e$.  In Sections
\ref{Section:HessianAndItsProperties} and
\ref{Section:Positivity}, we establish spectral properties and a
lower bound of the Hessian $\ActionAtSoliton''$ on the space
$\lb\dx^{-1}\TcaM\rb^{\bot}$.

The proof that $\HsNorm{1}{\e}$ is sufficiently small is the final
ingredient in the proof of the main theorem.  The remaining
sections concentrate on proving this crucial result.  We employ a
Lyapunov method and in Section \ref{Section:LyapDeriv} we
construct the Lyapunov function $\LyapunovFunctional$ and prove an
estimate on its time derivative. This estimate is later time
maximized over an interval $[0,T]$, and integrated to obtain an
upper bound on $\LyapunovFunctional$ involving the time $T$ and
the norms of $\e$. We combine this upper bound with the lower
bound on $\LyapunovFunctional$ following from the results of
Section \ref{Section:Positivity}, and obtain an inequality
involving $\HsNorm{1}{\e}$.  In Section \ref{Section:BoundOnFluct}
we solve the inequality to find an upper bound on $\HsNorm{1}{\e}$
provided $\HsNorm{1}{\e(0)}$ is small enough. We substitute this
bound into the bound appearing in the dynamical equation for $a$
and $c$, and take $\av\ev$ and $\epsilon_0$ small enough so that
all intermediate results hold to complete the proof.
\end{proof}

\section{Modulation of Solutions}
\label{Section:Decomposition} As stated in the previous section,
we begin the proof by decomposing the solution of
\eqref{Eqn:KdvGeneralizedWithPotential} into a modulated solitary
wave and a fluctuation $\e$:
\begin{eqnarray}
u(x,t)=Q_{c(t)a(t)}(x)+\e(x,t),
\label{EquationWithUErrorQDecomposition}
\end{eqnarray}
with $a$, $c$, and $\e$ fixed by the orthogonality condition
\begin{align}
\e\bot \SympOp\TcaM, \label{Cond:Orthogonality}
\end{align}
where
\begin{align*}
\SympOp:g\mapsto \int_{-\infty}^x g(z)\, dz.
\end{align*}
Note that $\SympOp\TcaM$ is a subset of $\Lp{2}$ (see
\eqref{Eqn:LIIAssumptions}).

The existence and uniqueness of parameters $a$ and $c$ such that
$\e=u-\Qca$ satisfies \eqref{Cond:Orthogonality} follows from the
next lemma concerning a restriction of $\SympOp$ and the implicit
function theorem.

The restriction $K$ of $\SympOp$ to the tangent space $\TcaM$ is
defined by the equation $K \TMProj=\TMProj\SympOp \TMProj$, where
$\TMProj$ is the orthogonal projection onto $\TcaM$.  In the
natural basis $\{\va,\vc\}$ of the tangent space $\TcaM$, the
matrix representation of $K$ is $N^{-1}\RestSympOp$, where
\begin{eqnarray*}
N&:=\lb\begin{array}{cc}
  \LpNorm{2}{\va}^2 & 0 \\
  0 & \LpNorm{2}{\vc}^2
\end{array}\rb
\end{eqnarray*}
and
\begin{eqnarray}
 \RestSympOp&:=
\lb\begin{array}{cc}
  \ip{\va}{\SympOp\va} & \ip{\vc}{\SympOp\va} \\
  \ip{\va}{\SympOp\vc} & \ip{\vc}{\SympOp\vc}
\end{array}\rb.
\label{Eqn:DefinitionOfOiN}
\end{eqnarray}
Recall that $\delta(c)=\frac{1}{2}\LpNorm{2}{\Qc}^2$.
\begin{lemma}
\label{Lemma:InvertibilityOfOiOmegaBound} If $\delta'(c)> 0$ on
the compact set $I\subset\Rp$, then the matrix $\RestSympOp$ is
invertible for all $c\in I$, and
\begin{align} \RestSympOpInv=\frac{1}{\delta'(c)}\lb\begin{array}{cc}
0 & 1\\
-1 & 0\\
\end{array}\rb.
\label{Eqn:LeadingOrderExpressionOfSymplecticInverse}
\end{align}
Clearly, $\|\RestSympOpInv\|\le \InfI{\delta'}^{-1}$, where
$\InfI{\delta'}:=\inf_I \delta'(c)$.
\end{lemma}
\begin{proof}
The lemma follows from the relations $\ip{\va}{\SympOp\va}=0$,
$\ip{\vc}{\SympOp\vc}=0$ and
$\ip{\va}{\SympOp\vc}=\ip{\vc}{\Qc}=\delta'(c)$.
\end{proof}

Given $\varepsilon>0$, define the tubular neighbourhood
$U_{\varepsilon}:=\{u\in\Lp{2}\,|\,\inf_{(c,\,a)\in
I\times\R}\LpNorm{2}{u-\Qca}<\varepsilon\}$ of the solitary wave
manifold $M_s$ in $\Lp{2}$.
\begin{prop}
\label{Prop:ExistenceOfDecomposition} Let $I\subset\Rp$ be a
compact interval such that $c\mapsto\Qca$ is $C^1(I)$.  Then there
exists a positive number
$\varepsilon_0=\varepsilon_0(I)=\O{\InfI{\delta'}^2}$ dependent on
$I$ and unique $C^1$ functions $
a:U_{\varepsilon_0}\rightarrow\Rp$ and
$c:U_{\varepsilon_0}\rightarrow I$, such that
\begin{equation*}
\ip{Q_{c(u)a(u)}-u}{\SympOp\zeta^{tr}_{c(u)a(u)}}=0\ \mbox{and}\
\ip{Q_{c(u)a(u)}-u}{\SympOp\zeta_{c(u)a(u)}^n}=0
\end{equation*}
for all $u\in U_{\varepsilon_0}$.  Moreover, there is a positive
real number $C=C(I)$ such that
\begin{equation}
\HsNorm{1}{u-Q_{c(u) a(u)}}\le C\inf_{\Qca\in
M_s}\HsNorm{1}{u-\Qca} \label{Ineq:InitialConditionIFT}
\end{equation}
for all $u\in U_{\varepsilon_0}\cap\Hs{1}$.
\end{prop}
\begin{proof}
Let $\mu:=(\mu^1,\mu^2)^T\in \Rp\times I$ and define $G:\Rp\times
I\times\Hs{1}\rightarrow\R^2$ as
\begin{eqnarray*}
G:(\mu,u)\mapsto\lb\begin{array}{c}
  \ip{\Qca-u}{\RestSympOp\va}\\
  \ip{\Qca-u}{\RestSympOp\vc}
\end{array}\rb,
\end{eqnarray*}
where $a=\mu^1$ and $c=\mu^2$.  The proposition is equivalent to
solving $G(g(u),u)=0$ for a $C^1$ function $g$.  Let $\mu_0=(a\,
c)^T$.  If $G$ is $C^1$, $G(\mu_0,\Qca)$=0, and $\p_\mu
F(\mu_0,\Qca)$ is invertible, then the implicit function theorem
asserts the existence of an open ball $B_{\varepsilon_0}(\Qca)$ of
radius $\varepsilon_0$ with centre $\Qca$, and a unique function
$g_{\Qca}:B_\delta(\Qca)\rightarrow\Rp\times I$, such that
$G(g_{\Qca}(u),u)=0$ for all $u\in B_{\varepsilon_0}(\Qca)$.  The
first two conditions are trivial, and the third follows from Lemma
\ref{Lemma:InvertibilityOfOiOmegaBound} since $\p_\mu
G(\mu_0,\Qca)=\RestSympOp$.  The radius of the balls
$B_\varepsilon(\Qca)$ depend on the parameters $c$ and $a$. To
obtain an estimate of the radius, and to show that we can take
$\varepsilon$ independent of the parameters $c$ and $a$, we give a
proof of the existence of the above function $g_{\Qca}$ for our
special case using the contraction mapping principle.

We wish to solve $G(\mu,u)=0$ for $\mu:=(\mu^1,\mu^2)^T$ with $u$
close to $\Qca$ in $\Lp{2}$.  Expand $G(\mu,u)$ in $\mu$ about
$\mu_0=(a\, c)^T$: $G(\mu,u)=G(\mu_0,u)+\p_\mu
G(\mu_0,u)(\mu-\mu_0)+R(\mu,u)$, with
$R(\mu,u)=\O{\|\mu-\mu_0\|^2}$ ($G$ is $C^2)$.  Thus, we must
solve $\mu=\mu_0-[\p_\mu G(\mu_0,u)]^{-1}\lb G(\mu_0,u)+R(\mu,u)
\rb$ for $\mu$.  Clearly, since $\p_\mu G(\mu_0,u)=\RestSympOp$,
$\mu$ must be a fixed point of
\begin{equation*}
H_{u \mu_0}(\mu):=\mu_0-\RestSympOpInv[G(\mu_0,u)+R(\mu,u)].
\end{equation*}
We now show that $H_{u \mu_0}$ is a strict contraction, and hence
has a fixed point.  By the mean value theorem
\begin{equation*}
\|H_{u \mu_0}(\mu_2)-H_{u \mu_0}(\mu_1)\|\le \sup\|\p_\mu H_{u
\mu_0}\|\|\mu_2-\mu_1\|,
\end{equation*}
where the supremum is taken over all allowed parameter values.
Furthermore, we have
\begin{align*}
\p_\mu H_{u \mu_0}(\mu)&=-\RestSympOpInv[\p_\mu G(\mu,u)-\p_u
G(\mu_0,u)]\\
&=-\RestSympOpInv[\p_\mu G(\mu,u)-\p_\mu G(\mu,\Qca)+\p_\mu
G(\mu,\Qca)-\p_\mu G(\mu_0,\Qca)+\p_\mu G(\mu_0,\Qca)-\p_u
G(\mu_0,u)]
\end{align*}
Using the mean value theorem again, we compute that
\begin{equation*}
\|\p_\mu G(\mu,u)-\p_\mu G(\mu_0, u)\|\le C_1\delta+
C_2\varepsilon
\end{equation*}
for some constants $C_1$ and $C_2$ if $\|\mu-\mu_0\|<\delta$ and
$\LpNorm{2}{u-\Qca}<\varepsilon$.  Combining all the estimates
gives
\begin{equation*}
\|H_{u \mu_0}(\mu_2)-H_{u \mu_0}(\mu_1)\|\le
\sup\|\RestSympOpInv\|\lb C_1\delta+C_2\varepsilon
\rb\|\mu_2-\mu_1\|.
\end{equation*}
Thus, if $\delta=\frac{1}{4}(C_1 \sup\|\RestSympOpInv\|)^{-1}$ and
$\varepsilon=\frac{1}{4}(C_2 \sup\|\RestSympOpInv\|)^{-1}$, then
$H_{u \mu_0}$ is a contraction.

We now choose $\delta$ and $\varepsilon$ so that $H_{u \mu_0}$
maps $B_\delta(\mu_0)$ to $B_\delta(\mu_0)$. We have that
\begin{equation*}
\|H_{u \mu_0}-\mu_0\|\le \|\RestSympOpInv \lb G(\mu_0, u)+R(\mu,u)
\rb\|\le \sup\|\RestSympOpInv\|\lb
\|G(\mu_0,u)-G(\mu_0,\Qca)\|+\O{\delta^2} \rb.
\end{equation*}
By the mean value theorem $\|G(\mu_0,u)-G(\mu_0,\Qca)\|\le
C_3\varepsilon$.  Thus, if we take
$\delta=\O{\sup\|\RestSympOpInv\|^{-1}}$ so that $\O{\delta^2}\le
\frac{1}{4}\lb\sup\|\RestSympOpInv\|\rb^{-1}\delta$, then
\begin{equation*}
\|H_{u \mu_0}-\mu_0\|\le
C_3\sup\|\RestSympOpInv\|\varepsilon+\frac{1}{4}\delta.
\end{equation*}
We now take $\varepsilon<\frac{1}{4}\lb C_3\sup\|\RestSympOpInv\|
\rb^{-1}\delta$ to obtain $\|H_{u \mu_0}-\mu_0\|\le
\frac{1}{2}\delta$.  To complete the argument, take $\delta$ to be
the smaller of $\frac{1}{4}\lb C_1\sup\|\RestSympOpInv\|\rb^{-1}$
and the above choice, and then $\varepsilon$ to be the smaller of
$\frac{1}{4}(C_2\sup\|\RestSympOpInv\|)^{-1}$ and $\delta(4
C_3\sup\|\RestSympOpInv\|)^{-1}$. Using the bound on
$\|\RestSympOpInv\|$ we find that
\begin{equation*}
\varepsilon=\O{\InfI{\delta'}^2}
\end{equation*}
if $\sup\|\RestSympOpInv\|\ge 1$, or equivalently, when
$\InfI{\delta'}$ is sufficiently small.

The above argument shows that there exists balls $\{
B_{\varepsilon}(\Qca)\, |\, a\in\Rp, c\in I\}$ with radius
$\varepsilon$ dependent only on the compact set $I$. Then,
defining $U_{\varepsilon_0}=\bigcup\{ B_{\varepsilon_0}(\Qca)\,
|\, a\in\Rp, c\in I\}$ and pasting the $C^1$ functions $g_{\Qca}$
together, into a $C^1$ function $g_{
I}:U_{\varepsilon_0}\rightarrow \Rp\times I$, proves existence of
the required $C^1$ functions $a(u)$ and $c(u)$. Uniqueness follows
from the uniqueness of the functions $g_{\Qca}$.

Let $u\in U_{\varepsilon}$, $c\in I$, and $a\in \R$, and consider
the equation
\begin{equation*}
u-Q_{c(u)a(u)}=u-\Qca+\Qca-Q_{c(u)a(u)}.
\end{equation*}
Clearly, inequality \eqref{Ineq:InitialConditionIFT} will follow
if $\HsNorm{1}{\Qca-Q_{c(u)a(u)}}\le C\HsNorm{1}{u-\Qca}$ for some
positive constant $C$.  Since the derivatives $\p_c\Qca$ and
$\p_a\Qca$ are uniformly bounded in $\Hs{1}$ over $I\times\R$, the
mean value theorem gives that $\HsNorm{1}{\Qca-Q_{c(u)a(u)}}\le
C\|(c,a)^T-(c(u),a(u))^T\|$, where the constant $C$ does not
depend on $c$, $a$.  The relations $g_{ I}(\Qca)=(c,a)^T$ and $g_{
I}(u)=(c(u),a(u))^T$ then imply $\HsNorm{1}{\Qca-Q_{c(u)a(u)}}\le
C\|g_{ I}(\Qca)-g_{ I}(u)\|$.  Again, we appeal to the mean value
theorem and use the properties of $\RestSympOp$ and that $\p_u g_{
I}=\p_\mu G^{-1}\p_u G$ is uniformly bounded in the parameters $c$
and $a$ to obtain \eqref{Ineq:InitialConditionIFT}.
\end{proof}

\section{Evolution Equations for Parameters $\e$, $a$ and $c$}
\label{Section:Projection} In Section \ref{Section:Decomposition}
we proved that if $u$ remains close enough to the solitary wave
manifold $M_s$, then we can write a solution $u$ to
\eqref{Eqn:KdvGeneralizedWithPotential} uniquely as a sum of a
modulated solitary wave $\Qca$ and a fluctuation $\e$ satisfying
the orthogonality condition \eqref{Cond:Orthogonality}. Thus, as
$u$ evolves according to the initial value problem
\eqref{Eqn:KdvGeneralizedWithPotential}, the parameters $a(t)$ and
$c(t)$ trace out a path in $\R^2$.  The goal of this section is to
derive the dynamical equations for the parameters $a$ and $c$, and
the fluctuation $\e$. We obtain such equations by substituting the
decomposition $u=\Qca+\e$ into
\eqref{Eqn:KdvGeneralizedWithPotential} and then projecting the
resulting equation onto appropriate directions, with the intent of
using the orthogonality condition on $\e$.

From now on, $u$ is the solution of
\eqref{Eqn:KdvGeneralizedWithPotential} with initial condition
$u_0$ satisfying $\epsilon_0:=\inf_{\Qca\in
M_s}\HsNorm{1}{u_0-\Qca}<\varepsilon_0$, and $T_0=T_0(u_0)$ is the
maximal time such that $u(t)\in U_\varepsilon$ for $0\le t\le
T_0$. Then for $0\le t\le T_0$, $u$ can be decomposed as in
\eqref{EquationWithUErrorQDecomposition} and
\eqref{Cond:Orthogonality}.

\begin{prop}
\label{Prop:EvolutionEquationAndBoundForAandC}
 Assume
$\delta'(c)\ne 0$.  Say $u=\Qca+\e$ is a solution to
(\ref{Eqn:KdvGeneralizedWithPotential}), where $\e$ satisfies
(\ref{Cond:Orthogonality}). Then, if $\HsNorm{1}{\e}$ is small
enough, $\ev\le 1$, and $c\in I$,
\begin{align}
\label{Eqn:DynamicalEquationForCAndA} \lb\begin{array}{c}
  \dot{a} \\
  \dot{c}
\end{array}\rb&=
\lb\begin{array}{c}
c-\B(t,a)\\
0
\end{array}\rb+
\B'(t,a)\frac{\delta(c)}{\delta'(c)}\lb\begin{array}{c}
0\\
1
\end{array}\rb+Z(a,\,c,\,\e),
\end{align}
where
$Z(a,\,c,\,\e)=\O{\av\ev^2+\av\ev\HsNorm{1}{\e}+\HsNorm{1}{\e}^2}$.

\end{prop}
\begin{proof}
Recall that the solitary wave $\Qca$ is an extremal of the
functional $\ActionAtSoliton$. To use this fact we rearrange
definition \eqref{DefinitionLS} of $\ActionAtSoliton$ to write the
Hamiltonian $\HamiltonianWithPotential$ as
\begin{equation*}
\HamiltonianWithPotential(u)=\ActionAtSoliton(u)-cP(u)+\frac{1}{2}\intR{\B
u^2(x)},
\end{equation*}
where for notational simplicity we have suppressed the space and
time dependency of $\B$. Substituting $\Qca+\e$ for $u$ in
\eqref{Eqn:KdVVariationalForm} and using the above expression for
$\HamiltonianWithPotential$ gives the equation
\begin{equation*}
\dot{a}\va+\dot{c}\vc+\dot{\e}=\SympOpInv\ActionAtSoliton'(\Qca+\e)-c\SympOpInv[\Qca+\e]+\SympOpInv[(\Qca+\e)\B],
\end{equation*}
where dots indicate time differentiation.  Let
$\Lca:=\ActionAtSoliton''(\Qca)$,
\begin{eqnarray*}
\DB:=\B(t,x)-\B(t,a)
\end{eqnarray*}
and
\begin{eqnarray*}
\RB:=\B(t,x)-\B(t,a)-\B'(t,a)(x-a).
\end{eqnarray*}
Taylor expanding $\ActionAtSoliton'(\Qca+\e)$ to linear order in
$\e$, using that $\Qca$ is an extremal of $\ActionAtSoliton$ and
the relation $\va=-\SympOpInv\Qca$ gives that
\begin{align}
\dot{\e}=\SympOpInv\left[(\Lca+\DB+\B(a)-c)\e\right]&+\SympOpInv\NpA{\e}-[\dot{a}-c+\B(a)]\va-\dot{c}\vc\nonumber\\&+\B'(a)\SympOpInv[(x-a)\Qca]+\SympOpInv[\RB\Qca].
\label{Eqn:KdVEquationForXiAndParameters}
\end{align}
The nonlinear terms have been collected into $\NpA{\e}$ given by
(\ref{Eqn:NpA}) in Appendix
\ref{Appendix:EstimateNonlinearRemainders}.

Define the vectors $\zeta_1:=\va$ and $\zeta_2:=\vc$.  Projecting
(\ref{Eqn:KdVEquationForXiAndParameters}) onto $\SympOp\zeta_i$
for $i=1,2$ and using the antisymmetry of $\SympOpInv$ gives the
two equations
\begin{align}
[\dot{a}-c+\B(a)]\left[
\ip{\va}{\SympOp\zeta_i}+\ip{\e}{\zeta_i}\right]&+\dot{c}\ip{\vc}{\SympOp\zeta_i}+\ip{\dot{\e}}{\SympOp\zeta_i}-\dot{a}\ip{\e}{\zeta_i}=-\B'(t,a)\ip{(x-a)\Qca}{\zeta_i}\nonumber\\
&-\ip{\RB\Qca}{\zeta_i}-\ip{\DB\e}{\zeta_i}-\ip{\NpA{\e}}{\zeta_i}-\ip{\Lca\e}{\zeta_i}.
\label{Eqn:KdVEquationForXiAndParametersSecond}
\end{align}
We can replace the term containing $\dot{\e}$ since the time
derivative of the orthogonality condition
$\ip{\e}{\SympOp\zeta_i}=0$ implies
$\ip{\dot{\e}}{\SympOp\zeta_i}=\dot{a}\ip{\e}{\zeta_i}-\dot{c}\ip{\e}{\dc\SympOp\zeta_i}$.
Note that we have used the relation $\p_a\zeta_i=-\dx\zeta_i$.
Thus, in matrix form,
(\ref{Eqn:KdVEquationForXiAndParametersSecond}) becomes
\begin{align}
(I+B)\RestSympOp\lb\begin{array}{c}\dot{a}-c+\B(t,a)\\
\dot{c}\end{array}\rb=X+Y, \label{Eqn:ApproximateDynamicalSystem}
\end{align}
where
\begin{align*}
X&:=-\B'(t,a)\delta(c)\lb\begin{array}{c}1\\0\end{array}\rb-\lb\begin{array}{c}\ip{\RB\Qca}{\va}\\ \ip{\RB\Qca}{\vc}\end{array}\rb,\\
Y&:=-\lb\begin{array}{c}\ip{\DB\e}{\va}+\ip{\NpA{\e}}{\va}+\ip{\Lca\e}{\va}\\
\\
\ip{\DB\e}{\vc}+\ip{\NpA{\e}}{\vc}+\ip{\Lca\e}{\vc}\end{array}\rb,
\end{align*}
and
\begin{eqnarray*}
B:=\lb\begin{array}{cc}\ip{\e}{\va} & \ip{\e}{\vc}\\\ip{\e}{\vc} &
-\ip{\e}{\dc\SympOp\vc}\end{array}\rb\RestSympOpInv.
\end{eqnarray*}
We have explicitly computed $\ip{(x-a)\Qca}{\zeta_i}$ to obtain
the above expression for $X$.

We now estimate the error terms and solve for $\dot{a}$ and
$\dot{c}$.  The assumption on the potential implies the bounds
\begin{align}
|\DB|\le \av\ev (x-a)\ \mbox{and}\ |\RB|\le \av\ev^2
(x-a)^2.\label{Eqn:SizeDV}
\end{align}
Thus, H\"{o}lder's inequality and exponential decay of $\Qca$
imply
\begin{align}
X&=-\B'(t,a)\delta(c) \lb\begin{array}{c}
1\label{Eqn:EstimateOnXFirst}\\
0
\end{array}\rb+\O{\av\ev^2}\\
&=\O{\av\ev}. \nonumber
\end{align}
Similarly, exponential decay of $\va$ and $\vc$ implies
$\ip{\DB\e}{\zeta_i}=\O{\av\ev\HsNorm{1}{\e}}$.  The linear term
$\ip{\Lca\e}{\zeta_i}$ is zero since $\Lca\va=0$, $\Lca\vc=-\Qca$
and $\e\bot \SympOp\va=-\Qca$. Lastly, $\ip{\NpA{\e}}{\zeta_i}\le
C\HsNorm{1}{\e}^2$ by the first estimate in Lemma
\ref{Appendix:EstimateNonlinearRemainders}.\ref{Lemma:NonlinearEstimates}.
Combining the above estimates gives the bound
\begin{align*}
\|Y\|=\O{\av\ev\HsNorm{1}{\e}+\HsNorm{1}{\e}^2}.
\end{align*}
By the second inclusion of (\ref{Eqn:LIIAssumptions}),
$\dc\SympOp\vc\in \Lp{2}$.  H\"{o}lder's inequality then implies
$\|B\|=\O{\HsNorm{1}{\e}}$. Thus, if $\HsNorm{1}{\e}$ is
sufficiently small, say so that $\|B\|\le \frac{1}{2}$, then $I+B$
is invertible and $\|\lb I+B\rb^{-1}\|\le 2$. Acting on equation
(\ref{Eqn:ApproximateDynamicalSystem}) by
$(I+B)^{-1}=I-B(I+B)^{-1}$ and then $\RestSympOpInv$ gives the
equation
\begin{align*}\lb
\begin{array}{c}\dot{a}-c+V(a)\\ \dot{c}\end{array}\rb=\RestSympOpInv
[X+B(I-B)^{-1} X+(I-B)^{-1} Y].
\end{align*}
Using the above estimates of $\|B\|$, $\|(I-B)^{-1}\|$, $\|X\|$,
and $\|Y\|$ implies
\begin{align*}\lb
\begin{array}{c}\dot{a}-c+V(a)\\ \dot{c}\end{array}\rb=\RestSympOpInv
X+\O{\av\ev\HsNorm{1}{\e}+\HsNorm{1}{\e}^2}.
\end{align*}
Replacing $X$ by (\ref{Eqn:EstimateOnXFirst}) completes the proof.
\end{proof}

\section{The Lyapunov Functional}
\label{Section:LyapDeriv} In the last section we derived dynamical
equations for the modulation parameters.  These equations contain
the $\Hs{1}$ norm of the fluctuation.  In this section we begin to
prove a bound on $\e$.  Recall that the latter bound is needed to
ensure that $u$ remains close to the manifold of solitary waves
$M_s$ for long time.

We employ a Lyapunov argument with Lyapunov function
\begin{align}
\LyapunovFunctional(t):=\ActionAtSoliton(\Qca+\e)-\ActionAtSoliton(\Qca)+\B'(a)\ip{(x-a)\Qca}{\e}.
\label{equ:LSDiffDef}
\end{align}
Remark: if $f(u)=u^3$, the last term in the Lyapunov functional is
not needed; however, apart from computational complexity, there is
no disadvantage in using the above function for this special case
as well.
\begin{lemma}
\label{Lemma:AlmostConservationOfLyapunov} Say $u=\Qca+\e$ is a
solution to (\ref{Eqn:KdvGeneralizedWithPotential}), where $\e$
satisfies (\ref{Cond:Orthogonality}).  Say $\av\le 1$. If
$\delta'(c)>0$, and $\ev$ and $\HsNorm{1}{\e}$ are less than 1,
with $\HsNorm{1}{\e}$ small enough, then
\begin{align}
\dd{t}
\LyapunovFunctional(t)&=\O{\av^2\ev^3+\lb\av\ev\tv+\av\ev^2\rb\HsNorm{1}{\e}+\av\ev\HsNorm{1}{\e}^2+\HsNorm{1}{\e}^4}.
\label{TimeDerivativeLiapunovFunctional}
\end{align}
\end{lemma}

\begin{proof}
Suppressing explicit dependence on $x$ and $t$, we have by
definition
\begin{align*}
\ActionAtSoliton(u):=\HamiltonianWithPotential(u)-\frac{1}{2}\intR{
u^2 \B}+cP(u).
\end{align*}
Thus, relations (\ref{Eqn:ConservationHamiltonian}),
(\ref{Eqn:ConservationMomentum}) and
(\ref{Eqn:ConservationPotentialMomentum}) imply that the time
derivative of $\ActionAtSoliton$ along the solution $u$ is
\begin{align*}
\dd{t}\ActionAtSoliton(u)=\intR{\frac{1}{2}\dot{c} u^2
+\B'\left[\frac{1}{2}c u^2- u f(u)+\frac{3}{2}(\dx
u)^2+F(u)\right]+\B''\, u \dx u}.
\end{align*}
Substituting $\Qca+\e$ for $u$, manipulating the result using
antisymmetry of $\dx$, and collecting appropriate terms into
$\B'(a)\ip{\Lca\e}{\SympOpInv((x-a)\Qca)}$,
$\ip{\NpA{\e}}{\dx[\DB(\Qca+\e)]}$, and
$\ip{\ActionAtSoliton'(\Qca)}{\dx(\DB(\Qca+\e))}$ gives the
relation
\begin{align*}
\dd{t}[\ActionAtSoliton(\Qca+\e)-\ActionAtSoliton(\Qca)]=&\B'(a)\ip{\Lca\e}{\SympOpInv((x-a)\Qca)}+\dot{c}\ip{\Qca}{\e}+\ip{\Lca\e}{\dx\lb\RB\Qca\rb}+\dot{c}\frac{1}{2}\LpNorm{2}{\e}^2\\
&+c\frac{1}{2}\ip{\B'\e}{\e}+\frac{3}{2}\ip{\B'\dx\e}{\dx\e}-\ip{f'(\Qca)\e}{\dx(\DB\e)}\\
&+\ip{\NpA{\e}}{\dx[\DB(\Qca+\e)]}+\ip{\B''\e}{\dx\e}+\ip{\ActionAtSoliton'(\Qca)}{\dx[\DB(\Qca+\e)]}.
\end{align*}
The last term is zero because $\ActionAtSoliton'(\Qca)=0$ and
since $\e\bot\Qca$, the quantity $\dot{c}\ip{\e}{\Qca}$ is also
zero.  We use Lemma \ref{Lemma:NonlinearEstimates}, assumptions
(\ref{Eqn:AssumptionOnPotential}) on the potential, estimates
(\ref{Eqn:SizeDV}), and
\begin{align*}
|\delta \B'|&\le\av\ev^2 x\
\end{align*}
to estimate the size of the time derivative.  We also use that
$\Qca$, $\dx\Qca$, $\dx^2\Qca$ and $f'(\Qca)$ are exponentially
decaying.  When $\ev\le 1$, higher order terms like
$\ip{\B''\e}{\dx\e}$ are bounded above by lower order terms like
$\ip{\B'\e}{\e}$.  Similarly, if $\HsNorm{1}{\e}\le 1$, then
$\av\ev\HsNorm{1}{\e}^2\le\av\ev\HsNorm{1}{\e}$.  This procedure
gives the estimate
\begin{align*}
\dd{t}[\ActionAtSoliton(\Qca+\e)-\ActionAtSoliton(\Qca)]=&\B'(a)\ip{\e}{\Lca\SympOpInv((x-a)\Qca)}+\ip{\NpA{\e}}{\DB\dx\e}\\
&+\O{|\dot{c}|\HsNorm{1}{\e}^2+\av\ev^2\HsNorm{1}{\e}+\av\ev\HsNorm{1}{\e}^2}.
\end{align*}
Applying the chain rule to the integrand of
\begin{equation*}
\intR{\dx\left[\lb
F(\Qca+\e)-F(\Qca)-f(\Qca)\e-\frac{1}{2}f'(\Qca)\e^2\rb\DB\right]}=0
\end{equation*}
and using the definition of $\NpA{\e}$ gives that
\begin{align*}
\ip{\NpA{\e}}{\DB\dx\e}=&\ip{\NpA{\e}+\frac{1}{2}f''(\Qc)\e^2}{\DB\dx\Qc}\\
&-\intR{\lb
F(\Qca+\e)-F(\Qca)-f(\Qca)\e-\frac{1}{2}f'(\Qca)\e^2\rb \B'}.
\end{align*}
The second estimate and the proof of the third estimate of
 Lemma \ref{Lemma:NonlinearEstimates} of Appendix \ref{Appendix:EstimateNonlinearRemainders} then imply the bound
$\ip{\NpA{\e}}{\DB\dx\e}=\O{\av\ev\HsNorm{1}{\e}^3}$. Thus, since
$\av\ev\HsNorm{1}{\e}^3\le \av\ev\HsNorm{1}{\e}^2$ when
$\HsNorm{1}{\e}\le 1$, we have
\begin{multline}
\dd{t}[\ActionAtSoliton(\Qca+\e)-\ActionAtSoliton(\Qca)]=\B'(a)\ip{\e}{\Lca\SympOpInv((x-a)\Qca)}+\O{|\dot{c}|\HsNorm{1}{\e}^2+\av\ev^2\HsNorm{1}{\e}+\av\ev\HsNorm{1}{\e}^2}.
 \label{Eqn:PropAlmostLiapunovConservationDLS}
\end{multline}
When $f(u)=u^3$, $\ip{\e}{\Lca\SympOpInv((x-a)\Qca)}=0$ since
$\vc=\SympOpInv[(x-a)\Qca]$.  In this special case the above
estimate is sufficient for our purposes, but in general, we need
to use the corrected Lyapunov functional. When $\e\in
C(\R,\,\Hs{1})\cap C^1(\R,\, \Hs{-2})$, $\B'(a)\ip{\e}{(x-a)\Qca}$
is continuously differentiable with respect to time;
\begin{align*}
\dd{t}\left[ \B'(a)\ip{\e}{(x-a)\Qca}
\right]=&\dt\B'\ip{\e}{(x-a)\Qca}+\B'(a)\ip{\dot{\e}}{(x-a)\Qca}+\dot{c}\B'(a)\ip{\e}{(x-a)\vc}\\
&+\dot{a}\B'(a)\ip{\e}{(x-a)\va}+\dot{a}\B''(a)\ip{\e}{(x-a)\Qca},
\end{align*}
where $\ip{\e}{\Qca}=0$ has been used to simplify the derivative.
Substituting for $\dt\e$ using
(\ref{Eqn:KdVEquationForXiAndParameters}) gives
\begin{align*}
\dd{t}[\B'(a)\ip{\e}{(x-a)\Qca}]=&-\B'(a)\ip{\e}{\Lca\SympOpInv((x-a)\Qca)}-[\dot{a}-c+\B(a)]\B'(a)\frac{1}{2}\LpNorm{2}{\Qca}^2+\dt\B'\ip{\e}{(x-a)\Qca}\\
&+[\dot{a}-c+\B(a)]\B'(a)\ip{\dx\e}{(x-a)\Qca}+[\dot{a}-c+\B(a)] \B''(a)\ip{\e}{(x-a)\Qca}\\
&+\dot{c} \B'(a)\ip{\e}{(x-a)\vc}-\B'(a)\ip{\e}{\DB\dx((x-a)\Qca)}-\B'(a)\ip{\NpA{\e}}{\dx((x-a)\Qca)}\\
&-\B'(a)\ip{\RB\Qca}{\dx((x-a)\Qca)}+[c-\B(a)]\B''(a)\ip{\e}{(x-a)\Qca}.
\end{align*}
We estimate using the same assumptions used to derive
(\ref{Eqn:PropAlmostLiapunovConservationDLS}).  If
$\HsNorm{1}{\e}$ and $\ev$ are less than 1, then
\begin{align*}
\dd{t}[\B'(a)\ip{\e}{(x-a)\Qca}]=&-\B'(a)\ip{\e}{\Lca\SympOpInv((x-a)\Qca)}+\O{ |\dot{a}-c+\B(a)|\av\ev+|\dot{c}|\av\ev\HsNorm{1}{\e}}\\
&+\O{\av^2\ev^3+((1+\av)\ev^2+\ev\tv)\av\HsNorm{1}{\e}+\av\ev\HsNorm{1}{\e}^2}.
\end{align*}
Adding the above expression to
(\ref{Eqn:PropAlmostLiapunovConservationDLS}) gives an upper bound
containing $|\dot{c}|$ and $|\dot{a}-c+\B(a)|$. Replacing these
quantities using the bounds
\begin{equation*}
|\dot{c}|=\O{\av\ev+\av\ev\HsNorm{1}{\e}+\HsNorm{1}{\e}^2}\\
\end{equation*}
and
\begin{equation*}
|\dot{a}-c+\B(a)|=\O{\av\ev^2+\av\ev\HsNorm{1}{\e}+\HsNorm{1}{\e}^2}\\
\end{equation*}
from Proposition \ref{Prop:EvolutionEquationAndBoundForAandC}, and
bounding higher order terms by lower order terms gives
(\ref{TimeDerivativeLiapunovFunctional}). To use the above bounds
on $|\dot{c}|$ and $|\dot{a}-c+\B(a)|$ we must assume
$\HsNorm{1}{\e}$ is small enough so that Proposition
\ref{Prop:EvolutionEquationAndBoundForAandC} holds.
\end{proof}

\section{Spectral Properties of the Hessian $\Lca$}
\label{Section:HessianAndItsProperties} The Hessian
$\p^2\ActionAtSoliton$ at $\Qca$ in the $\Lp{2}$ pairing is
computed to be the unbounded operator
\begin{align}
\Lca&:=-\dx^2+c-f'(\Qca), \label{Eqn:Hessian}
\end{align}
defined on $\Lp{2}$ with domain $\Hs{2}$.  We extend this operator
to the corresponding complex spaces.
\begin{prop}  \label{Prop:Spectrum}
The self-adjoint operator $\Lca$ has the following properties.
{\begin{enumerate}
    \item $\Lca\va=0$ and $\Lca\vc=-\Qca$.
    \item All eigenvalues of $\Lca$ are simple, and
    $\Null{\Lca}=\Span{\va}$.
    \item $\Lca$ has exactly one negative eigenvalue.
    \item The essential spectrum is $[c,\infty)\subset\Rp$.
    \item $\Lca$ has a finite number of eigenvalues in $(-\infty,
    c)$.
\end{enumerate}}
\end{prop}
\begin{proof}
Recall that the vectors $\va:=-\dx\Qca$ and $\vc:=\dc\Qca$ are in
the Sobolev space $\Hs{2}$.  Thus, relations $\Lca\va=0$ and
$\Lca\vc=-\Qca$ make sense, and are obtained by differentiating
$\ActionAtSoliton'(\Qca)=0$ with respect to $a$ and $c$.  The
first relation above proves that $\va$ is a null vector.

Say $\zeta,\eta\in \Hs{2}$ are linearly independent eigenvectors
of $\Lca$ with the same eigenvalue. Then, since $\Lca$ is a second
order linear differential operator without a first order
derivative, the Wronskian
\begin{eqnarray*}
W(\eta,\zeta)=\zeta\dx \eta-\eta\dx\zeta
\end{eqnarray*}
is a non-zero constant.  With $\eta$ and $\zeta$ both in $\Hs{2}$
however, the limit $\lim_{x\rightarrow \infty} W(\eta,\zeta)$ is
zero. This contradicts the non vanishing of the Wronskian, and
hence all eigenvalues of $\Lca$ are simple and, in particular,
$\Null{\Lca}=\Span{\va}$.

Next we prove that the operator $\Lca$ has exactly one negative
eigenvalue using Sturm-Liouville theory on an infinite interval.
Recall that the solitary wave $\Qca(x)$ is a differentiable
function, symmetric about $x=a$ and monotonically decreasing if
$x>a$. This implies that the null vector $\va$, or equivalently,
the derivative of $\Qca$ with respect to $x$, has exactly one root
at $x=a$. Therefore, by Sturm-Liouville theory, zero is the second
eigenvalue and there is exactly one negative eigenvalue.

We use standard methods to compute the essential spectrum.  Since
the function $f'(\Qca(x))$ is continuous and decays to zero at
infinity, the bottom of the essential spectrum begins at
$\lim_{x\rightarrow \infty} (c-f'(\Qca(x)))=c$ and extends to
infinity: $\sigma_{ess}(\Lca)=[c,\infty)$. Furthermore, the bottom
of the essential spectrum is not an accumulation point of the
discrete spectrum since $f'(\Qca(x))$ decays faster than $x^{-2}$
at infinity.  Hence, there is at most a finite number of
eigenvalues in the interval $(-\infty,c)$.  For details see
\cite{ReSiI, ReSiIV, GuSi2003}.
\end{proof}

\section{Strict Positivity of the Hessian}
\label{Section:Positivity} In this section we prove strict
positivity of the Hessian $\Lca$ on the orthogonal complement to
the 2-dimensional space $\SympOp\TcaM=\Span{\Qca,\,\SympOp\vc}$.
This result is a crucial ingredient needed to prove the bound on
the fluctuation $\e$.
\begin{prop}
\label{Prop:Positivity} Assume $\delta'(c)>0$ on $I\subset\Rp$. If
$\e\bot\SympOp\TcaM$, then there is a positive constant $\rho$
such that $\ip{\Lca\e}{\e}\ge\rho\HsNorm{1}{\e}^2$.
\end{prop}

\begin{proof}
Define $X:=\{\e\in\Hs{1}\ |\ \e\bot\SympOp\TcaM,\
\LpNorm{2}{\e}=1\}$. By the max-min principle, $\inf_{X\cap\Hs{2}}
\ip{\Lca\e}{\e}$ is attained or is equal to $\inf
\sigma_{ess}(\Lca)=c$.  If the later holds the proof is complete.
In the former case, let $\eta$ be the minimizer.

We claim the set of vectors $\{\va,\vc,\eta\}$ is an linearly
independent set. If they were dependent, then, since $\va$ and
$\vc$ are orthogonal, there are non-zero constants $\alpha$ and
$\beta$ such that $\eta=\alpha\va+\beta\vc$.  Projecting this
equation onto $\SympOp\va$ and $\SympOp\vc$ gives the equations
$\beta\delta'(c)=0$ and $\alpha\delta'(c)=0$.  Thus, the
assumption $\delta'(c)>0$ implies $\eta=0$.  A contradiction since
the zero function does not lie in the set $X$.  Note that in
deriving $\alpha\delta'(c)=0$ we have used that $\SympOp$ is
antisymmetric on the span of $\vc$ since $\SympOp\vc\in \Lp{2}$.

By the min-max principle, if
\begin{align*}
\lambda_3&:=\inf \left\{ \max \left\{\ip{\Lca\e}{\e}\, |\, \e\in
V,\, \LpNorm{2}{\e}=1 \right\} \, |\, V\subset \Hs{2},\,
\mbox{dim}\,
V=3 \right\}\\
&\le \max \left\{ \ip{\Lca\e}{\e} \,|\,
\e\in\Span{\va,\,\vc,\,\eta}\right\}
\end{align*}
is below the essential spectrum, then it is the third eigenvalue
counting multiplicity.  Let $\e=\alpha\eta+\beta\va+\gamma\vc$
where $\alpha$, $\beta$ and $\gamma$ are arbitrary apart from
satisfying $\LpNorm{2}{\e}=1$. Thus, since the third eigenvalue of
$\Lca$ is positive (see Section
\ref{Section:HessianAndItsProperties}),
\begin{align*}
0<\ip{\Lca\e}{\e}=\alpha^2\ip{\Lca\eta}{\eta}-\gamma^2\delta'(c)\le\alpha^2\ip{\Lca\eta}{\eta},
\end{align*}
and hence $\ip{\Lca\eta}{\eta}>0$.  The function
$\sigma(c)=\ip{\Lca\eta}{\eta}$ is continuous since both
$\SympOp\va$ and $\SympOp\vc$ are continuous in $\Lp{2}$ as
functions of $c$. Set $\varrho=\inf_I \sigma(c)$.

We now improve the result to an $\Hs{1}$ norm.  If we define the
constant $K(I):=\sup_I \SupNorm{c-f'(\Qca)}$, then
$\ip{\Lca\e}{\e}\ge\LpNorm{2}{\dx\e}^2-K(I)\LpNorm{2}{\e}^2$.
Adding to this bound the factor $\frac{K+1}{\varrho}$ of the lower
bound $\ip{\Lca\e}{\e}\ge \varrho\LpNorm{2}{\e}^2$ derived above
completes the proof.
\end{proof}


\section{Bound on the Fluctuation}
\label{Section:BoundOnFluct} We are now ready to prove the bound
on the fluctuation.
\begin{prop}
\label{Prop:FluctuationBound} Say $\av\le 1$.  Then, for small
enough $\ev\le 1$ and initial fluctuation $\HsNorm{1}{\e(0)}\le
1$, there exists a constant $C$ such that the bound
\begin{align*}
\HsNorm{1}{\e(t)}=\O{\epsilon_0+\lb\av\ev\rb^\frac{1}{2}\epsilon_0^\frac{1}{2}+\ev+\tv}
\end{align*}
holds for all times $t\le T=C\lb\av\ev\rb^{-1}$.
\end{prop}
\begin{proof}
Lemma \ref{Lemma:AlmostConservationOfLyapunov} implies
\begin{align*}
\left|\dd{t} \LyapunovFunctional(t)\right|\le
C\lb\av^2\ev^3+\lb\av\ev\tv+\av\ev^2\rb\HTNorm{\e}+\av\ev\HTNorm{\e}^2+\HTNorm{\e}^4\rb
\end{align*}
for some constant $C>0$ where $\HTNorm{\e}:=\sup_{0\le t\le
T}\HsNorm{1}{\e}$. Integrating over $[0,T]$ gives an upper bound
on $\LyapunovFunctional(T)$.  A lower bound is obtained by
expanding $\ActionAtSoliton(\Qca+\e)$ to quadratic order then
using Proposition \ref{Prop:Positivity}, the third estimate of
Lemma \ref{Lemma:NonlinearEstimates} and
$V'(a)\ip{\e}{(x-a)\Qca}=\O{\av\ev\HsNorm{1}{\e}}$. We obtain,
after setting all non-essential constants to one,
\begin{align*}
\HTNorm{\e}^2-\HTNorm{\e}^3-\av\ev\HTNorm{\e}\le
\LyapunovFunctional(T)\le
|\LyapunovFunctional(0)|+\lb\av^2\ev^3+\lb\av\ev\tv+\av\ev^2\rb\HTNorm{\e}+\av\ev\HTNorm{\e}^2+\HTNorm{\e}^4\rb
T
\end{align*}
for all $T>0$.  Take $T=\O{\lb\av\ev\rb^{-1}}$.  Then, under the
smallness assumption $\HsNorm{1}{\e}\ll(\av\ev)^\frac{1}{2}$,
\begin{align*}
\HsNorm{1}{\e}=\O{|\LyapunovFunctional(0)|^\frac{1}{2}+\ev+\tv}.
\end{align*}

The initial value of the Lyapunov functional
$\LyapunovFunctional(0)$ can be bounded by the $\Hs{1}$ norm of
the initial fluctuation $\HsNorm{1}{\e(0)}\le C\epsilon_0$ (recall
that $\epsilon_0:=\inf_{\Qca\in M_s}\HsNorm{1}{u_0-\Qca}$. Indeed,
Taylor expanding $\ActionAtSoliton(\Qca+\e)$ to second order in
$\e$ and using the third estimate in Lemma
\ref{Lemma:NonlinearEstimates} gives
$|\LyapunovFunctional(0)|=\O{\epsilon_0^2+\av\ev\epsilon_0}$ if
$\epsilon_0\ll1$.  To complete the proof we take $\ev$ and
$\epsilon_0$ small enough so that $\HsNorm{1}{\e(t)}$ is
sufficiently small for Lemma
\ref{Lemma:AlmostConservationOfLyapunov} to hold.
\end{proof}

We now prove the main theorem.
\begin{proof}[Proof of Theorem \ref{MainThm}]
By our choice $\epsilon_0<\varepsilon_0$, there is a (maximal)
time $T_0$ such that the solution $u$ in
\eqref{Eqn:KdvGeneralizedWithPotential} is in $U_{\varepsilon_0}$
for time $t\le T_0$.  Hence decomposition
\eqref{EquationWithUErrorQDecomposition} with
\eqref{Cond:Orthogonality}, and Proposition
\ref{Prop:FluctuationBound} are valid for the solution $u$ over
this time and imply the statements of the main theorem.  In
particular $\HsNorm{1}{\e(t)}=
\O{\epsilon_0+\lb\av\ev\rb^\frac{1}{2}\epsilon_0^\frac{1}{2}+\ev+\tv}$
for times $t\le \min\{T_0, T\}$.  Taking
$\epsilon_0+\lb\av\ev\rb^\frac{1}{2}\epsilon_0^\frac{1}{2}+\ev+\tv\ll\varepsilon_0$,
we must have $t\le T$ by maximality of the time $T_0$.
\end{proof}

\appendix
\section{Estimates of Nonlinear Remainders}
\label{Appendix:EstimateNonlinearRemainders} \indent\indent Define
\begin{align*}
\NA{\e}:=-\intR{F(\Qc+\e)-F(\Qc)-F'(\Qc)\e-\frac{1}{2}F''(\Qc)\e^2}
\end{align*}
and
\begin{align}
\NpA{\e}:=-\lb f(\Qc+\e)-f(\Qc)-f'(\Qc)\e\rb. \label{Eqn:NpA}
\end{align}
Note that $\NpA{\e}=\p_\xi\NA{\e}$ under the $\Lp{2}$ pairing.
\begin{lemma}
\label{Lemma:NonlinearEstimates} If $\HsNorm{1}{\e}\le 1$ and
$f\in C^k(\R)$ for some $k\ge 3$, with $f^{(k)}\in\Lp{\infty}$,
then there are positive constant $C_1$, $C_2$, and $C_3$ such that
\begin{enumerate}
    \item $\LpNorm{2}{\NpA{\e}}\le C_1\HsNorm{1}{\e}^2$\label{Item:NonlinearEstimateNpA}
    \item $\LpNorm{2}{\NpA{\e}+\frac{1}{2}f''(\Qca)\e^2}\le
    C_2\HsNorm{1}{\e}^3$\label{Item:NonlinearEstimateNpAMore}
    \item $\left|\NA{\e}\right|\le C_3\HsNorm{1}{\e}^3$.\label{Item:NonlinearEstimateNA}
\end{enumerate}
\end{lemma}
\begin{proof}
Taylor's remainder theorem implies
\begin{align*}
\NpA{\e}=-\sum_{n=2}^{k-1}\frac{1}{n!}f^{(n)}(\Qca)\e^n-R(\Qca,
\e),
\end{align*}
where, since $f^{(k)}\in\Lp{\infty}$, $|R(\Qca,\e)|\le C|\e|^k$.
Recall that $\Qca$ is continuous and decays exponentially to zero.
Together with the assumption that $f\in C^k(\R)$, this implies
$f^{(n)}(\Qca)\in\Lp{\infty}$ for $2\le n\le k-1$.  Thus, after
pulling out the largest constant,
\begin{align*}
\LpNorm{2}{\NpA{\e}}\le C\sum_{n=2}^k\LpNorm{2}{\e^n}.
\end{align*}
To obtain item \ref{Item:NonlinearEstimateNpA} we use the bound
$\LpNorm{2}{\e^n}\le C\HsNorm{1}{\e}^n$, which is obtained from
the inequality $\LpNorm{\infty}{\e}\le C\HsNorm{1}{\e}$ and the
assumption that $\HsNorm{1}{\e}\le 1$.

Clearly, slight modification of the above proof gives items
\ref{Item:NonlinearEstimateNpAMore} and
\ref{Item:NonlinearEstimateNA}.  For the latter we use that the
assumptions on $f$ imply $F\in C^{k+1}(\R)$ with
$F^{(k+1)}\in\Lp{\infty}$.
\end{proof}

\def\cydot{\leavevmode\raise.4ex\hbox{.}} \def\cprime{$'$}

\end{document}